\documentclass[10pt, twocolumn]{article}
\pdfoutput=1
\usepackage{graphicx} 
\usepackage[inkscapearea=page]{svg}
\usepackage{subcaption}
\usepackage[labelfont=bf]{caption}
\usepackage[
    backend=biber,
    style=numeric-comp,
    sorting=none,
    date=year,
    isbn=false,
    url=false,
    maxbibnames=4, 
]{biblatex}
\usepackage{hyperref}
\usepackage{comment}
\usepackage{physics}
\usepackage{lipsum}
\usepackage{titlesec}
\titleformat*{\section}{\large\bfseries}
\usepackage{geometry}
\usepackage{authblk}
\usepackage{tablefootnote}

\title{Ultra-fast Digital DPC Yielding High Spatio-Temporal Resolution for Low-Dose Phase Characterisation}

\author[1,2,*]{Julie Marie Bekkevold}
\author[1,2]{Jonathan J. P. Peters}
\author[3]{Ryo Ishikawa}
\author[3]{Naoya Shibata}
\author[1,2]{Lewys Jones}
\affil[1]{School of Physics, Trinity College Dublin, College Green, Dublin D02 PN40, Ireland}
\affil[2]{Advanced Microscopy Laboratory, Centre for Research on Adaptive Nanostructures and Nanodevices (CRANN), Trinity College Dublin, Dublin D02 DA31, Ireland}
\affil[3]{Institute of Engineering Innovation, University of Tokyo, Bunkyo, Tokyo 113-8656, Japan}

\date{August 16, 2024}

\begin{document}

\makeatletter
\twocolumn[
  \maketitle
  *Corresponding author: \href{mailto:bekkevoj@tcd.ie}{bekkevoj@tcd.ie}\\
    \begin{@twocolumnfalse}
        \begin{abstract}
            In the scanning transmission electron microscope, both phase imaging of beam-sensitive materials and characterisation of a material's functional properties using \textit{in-situ} experiments are becoming more widely available. 
            As the practicable scan speed of 4D-STEM detectors improves, so too does the temporal resolution achievable for both differential phase contrast (DPC) and ptychography. 
            However, the read-out burden of pixelated detectors, and the size of the gigabyte to terabyte sized data sets, remain a challenge for both temporal resolution and their practical adoption. 
            In this work, we combine ultra-fast scan coils and detector signal digitisation to show that a high-fidelity DPC phase reconstruction can be achieved from an annular segmented detector. 
            Unlike conventional analog data phase reconstructions from digitised DPC-segment images yield reliable data, even at the fastest scan speeds. 
            Finally, dose fractionation by fast scanning and multi-framing allows for post-process binning of frame streams to balance signal-to-noise ratio and  temporal resolution for low-dose phase imaging for \textit{in-situ} experiments. 
        \end{abstract}
    \end{@twocolumnfalse}

    \vspace{3ex}
]

\section{Introduction}
Scanning transmission electron microscopy (STEM) is a commonly used technique for the structural, functional and chemical characterisation of materials at very high spatial resolution.
In recent decades, the traditional monolithic bright field (BF) and annular dark field (ADF) detectors have started to be superseded by other geometries of detectors enabling new imaging techniques. 
One of these is the development of high dynamic range cameras capable of capturing the entire two-dimensional (2D) convergent beam electron diffraction (CBED) pattern for every 2D scan point on the specimen. 
This results in a four-dimensional (4D) dataset, and the detectors are therefore aptly referred to as 4D-STEM or pixelated detectors. 
These 4D detectors are becoming increasingly popular for many different types of materials characterisation with imaging methods that require position-sensitive detectors to elucidate the internal structure of the bright field disk and overlapping regions with scattered disks.
\par 

A higher pixelation of the detector allows for a finer sampling of the CBED patterns at each probe position.
However, the greater the number of pixels, the higher the read-out overhead from the detector, and the slower the minimum sample-plane dwell time achievable. 
When using pixelated detectors the minimum dwell time, or the minimum fly-back time, are not limited by the scan coils and generators of the microscope but rather by the detector read-out.
The necessity of these increased dwell times in 4D-STEM can then introduce unwanted image artefacts due to sample drift or environmental instability~\cite{jones_identifying_2013,oleary_increasing_2022,grieb_gan_2024}, and makes \textit{in-situ} characterisation challenging.
\par 
\begin{figure*}[htb]
    \centering
    \includegraphics[width=.9\textwidth]{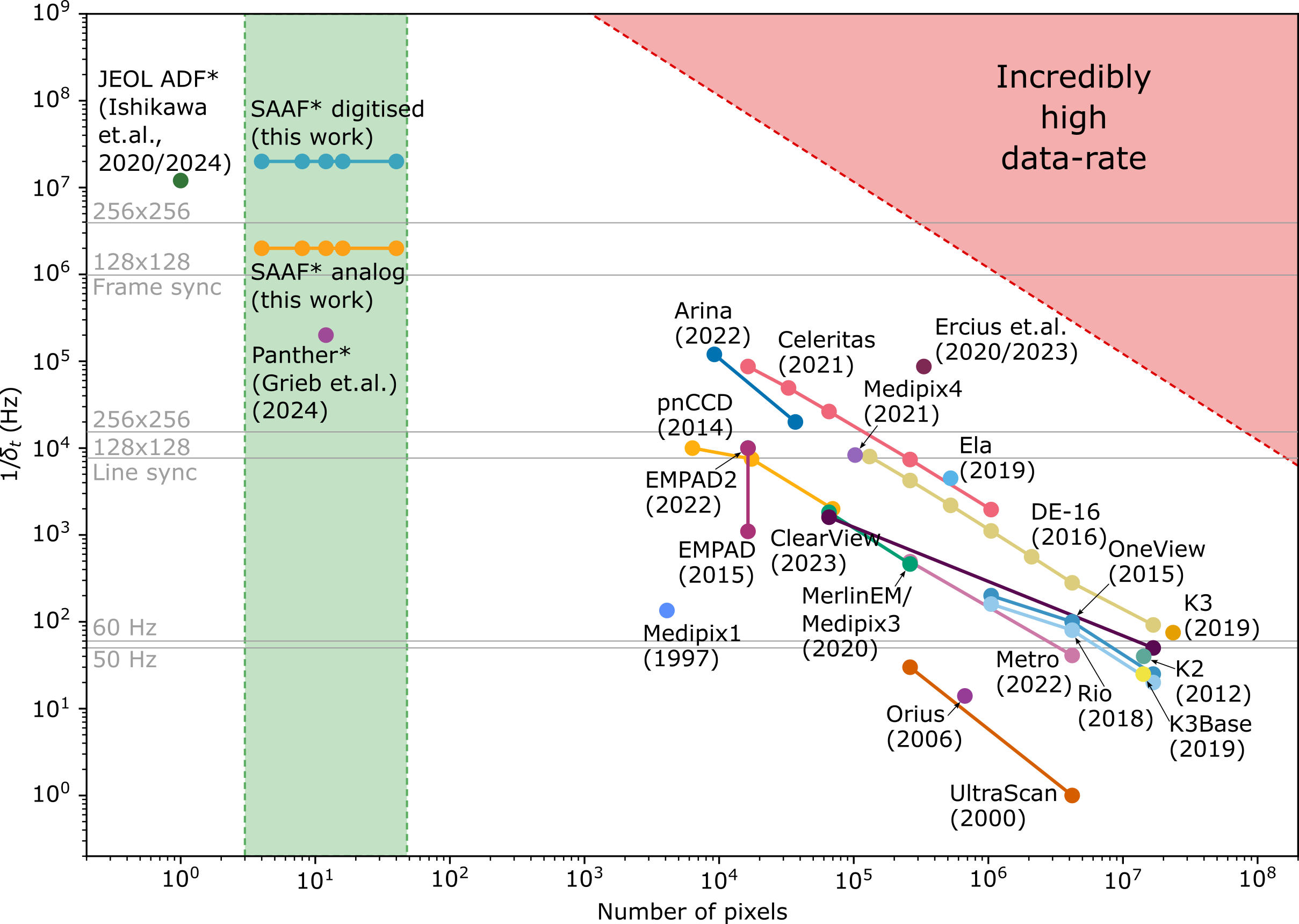}
    \caption{Illustration of the practicable scan speed (in $\frac{1}{\delta_t}$) versus number of segments/pixels for various commercially available or researched detectors. 
    Several detectors may be operated in different binning modes, where higher binning allows for faster data readout due to lower pixel numbers. 
    All the data for the 4D detectors has been acquired from data-sheets or publications reporting the highest practicable readout speed from the detectors~\cite{bardelloni_new_2000,ponchut_maxipix_2011,ballabriga_medipix3_2011,tate_high_2016,ishikawa_high_2020,ballabriga_introduction_2020,wang_electron_2021,philipp_very-high_2022,ercius_4d_2023}. For the detectors marked with * the scan speed is taken from publications and does not necessarily represent the highest possible speed achievable~\cite{grieb_gan_2024,ishikawa_real-time_2024}. Where several bit depth modes are available, the main reported readout speed is quoted (for details see Supplementary Information). 
    Data used may be found at~\cite{detector-frame-rate-data}.}
    \label{fig:detectors_fps_vs_pixels}
\end{figure*}
Recent advances in ptychographic reconstruction reaching information limits of less than $0.5$~Å uses both a high dwell time and high beam current which is challenging for samples that damage under beam irradiation~\cite{nguyen_achieving_2024}. 
The core assumption for ptychographic reconstruction that the sample is static poses challenges if the goal is to image dynamic processes \textit{in-situ}. 
Especially when the minimum practicable dwell time for pixelated detectors is still typically found in the $10-1000$~\textmu s regime. 
The assumption of the static object phase would only hold for samples that vary on time scales much higher than a few seconds, depending on the number of scan points (field of view) and the resolution required. 
Still, it should be mentioned that with iterative ptychography and defocused probes the dose delivered to the sample may be reduced and temporal resolution improved due to the increased sample-plane step size~\cite{rodenburg_phase_2004,maiden_improved_2009,zhou_low-dose_2020}. 
\par

Since pixelated detectors were first used to acquire images of the CBED in a STEM scan in the late 90's with the first Medipix-detector, the data rate has been increasing with the development of faster and higher pixel-number detectors; these include the TFS EMPAD (2015), Quantum Detectors MerlinEM (Medipix3, 2020), DirectElectron Celeritas (2021), and Dectris Arina (2022).
To contextualise the speed and pixelation of various detectors, these parameters are plotted in \autoref{fig:detectors_fps_vs_pixels}. 
Most pixelated detectors support hardware binning of pixels to get a higher read-out speed at the expense of a lower CBED sampling. 
This binning leads to a lower read-out delay and data set size, and facilitates higher real-space scan speeds and for this reason some detectors in \autoref{fig:detectors_fps_vs_pixels} are represented by multiple points.
Further, the type and dynamic ranges of the commercially available 4D-STEM detectors in \autoref{fig:detectors_fps_vs_pixels} are summarised in Table S1 in the Supplementary Information. 
For several detectors the frame-rate may be increased by choosing a lower bit depth, essentially sacrificing dynamic range for speed. 
Still, bit depth does not tell the full story of the dynamic range achievable by the detector since some detectors encode more information than intensity recorded in the pixel in their output, with the result that a higher dynamic range is supported.
At the time of writing, in terms of frame-rate for fully pixelated 4D detectors, the state of the art is the Dectris Arina detector with a minimum real-space dwell time of $10$~\textmu s.~\cite{stroppa_stem_2023}.

Notice how, beginning with the Medipix 1 launched in 1997, the community has generally moved towards higher numbers of pixels and increasing practicable scan speed, approaching the generation of multi-TB datasets in hours -- if not minutes~\cite{spurgeon_towards_2021,oleary_increasing_2022}.
For completeness, a `single element' detector case of ADF STEM is shown in the upper-left of \autoref{fig:detectors_fps_vs_pixels}.  We use the term elements here as a general term that can mean either pixels or detection segments. Segmented detectors from 4 to 40 elements fall within the green highlighted region.
There is then a `pixelation gap' in the range from 40 to \(\sim \)3000 elements which the authors imagine may be an active area of technology development in the coming years.
\par 

To avoid image distortions arising from  power supply instability, it is often desirable to sync the start of each pixel, line or even frame to the power supply frequency~\cite{jones_identifying_2013}. 
In \autoref{fig:detectors_fps_vs_pixels}, the horisontal lines labelled 50 and 60 Hz indicate the threshold of dwell time (camera frame rate) where syncing the start of each pixel to the power supply frequency is possible. For example, this shows that the Gatan UltraScan CCD, released around the year 2000, is not able to pixel-sync to A.C. mains frequency for 4D-STEM. 
Similarly, the horizontal lines labelled ``Line sync" indicate the minimum frame speed (maximum dwell time) where syncing the start of each line to a 60 Hz mains frequency becomes possible, in a 128 by 128 or 256 by 256 pixel image respectively.
Lastly, the ``Frame sync" lines indicate the scan speed at which syncing the start of each frame to a 60 Hz mains frequency would be possible, again for 128 by 128 and 256 by 256 pixel images. 
\par 

With the 4D nature of the data sets, the increasing number of pixels in the detectors, and the hardware's ever-increasing scan speed, experiment data rate increases massively. 
We are now in the era where generating massive (multiple PB) data sets from single microscope sessions is no longer uncommon~\cite{spurgeon_towards_2021,oleary_increasing_2022}. 
The triangular area in the top right corner of \autoref{fig:detectors_fps_vs_pixels} corresponds roughly to the range where Terabit per second data-rates (Tbps) would start to arise, but the question remains whether these massive data sets are truly necessary.
From a sustainability perspective, these high data rates are problematic. 
Abiding by the FAIR data principles, requiring our data to be Findable, Accessible, Interoperable, and Reusable, demands that we store all our raw data in non-proprietary data formats, in an accessible way~\cite{wilkinson_fair_2016,barker_introducing_2022}. 
Most research data is required to stay stored and available for at least a few years requiring on-site hard-drives or cloud storage space. 
This often means that cloud storage, and online repositories such as Zenodo~\cite{https://doi.org/10.25495/7gxk-rd71}, are the only viable options regardless of the carbon footprint of that.
Electricity consumed by data centres currently accounts for $0.3$~\% of world carbon emissions, and storing 1TB of data in the cloud results in 2 tons of CO$_2$ emissions per year~\cite{weber_energy_2010,jones_how_2018,monserrate_cloud_2022}. 
For comparison, this is about the same carbon footprint as a round-trip flight from Dublin to San Francisco~\cite{ciers_carbon_2019}.
\par

While 4D-STEM enables high-resolution imaging in both the image and detector plane, there are also position-sensitive detectors with a lower number of discrete segments which have been shown to yield high-quality image data for phase contrast techniques such as ptychography, optimum bright field (OBF), and differential phase contrast (DPC).
Specifically, DPC has long been established as a powerful tool for probing both magnetic and electric fields inside samples, including the local electric fields surrounding individual atoms~\cite{rose_phase-contrast_1974,dekkers_differential_1974,lohr_differential_2012,shibata_differential_2012}.
Although reliable DPC signals depend on minimal dynamical scattering, and thus requires a very thin sample, there are several examples where the functional properties of specimens, such as ferroelectric domain formation or internal magnetic fields, have been characterised with fewer numbered segmented detectors~\cite{ooe_ultra-high_2021,shibata_differential_2012,chen_imaging_2020}.
Furthermore, a significant amount of noteworthy high-quality DPC imaging has been done with as low as four annular segments~\cite{shibata_new_2010,shibata_differential_2012,close_towards_2015,shibata_electric_2017,ishikawa_spatial_2023,grieb_gan_2024}.
Similarly, almost a decade ago, Yang et.~al. showed that for single-side-band (SSB) ptychography the signal-to-noise ratio (SNR) of the reconstructed image does not increase significantly with increased detector plane pixelation beyond around 16x16~\cite{yang_efficient_2015}. 
And just a few years ago, Ooe et.~al. demonstrated impressive low-dose phase reconstruction with the OBF technique with a 16 segmented annular detector~\cite{ooe_ultra-high_2021}. 
\par

For an annular, four-quadrant segmented detector, the minimum dwell time for precise scanning is no longer limited by the read-out from the detector but rather by the practicable scan speed of the STEM itself. 
Once the readout overhead becomes negligible, the STEM scan may be done at very high speeds with dwell times far less than a microsecond. 
With high-speed scan coils and state-of-the-art scan controllers, dwell times as low as $50$~ns are readily available~\cite{ishikawa_high_2020,ishikawa_spatial_2023}. 
This low dwell time enables multi-frame imaging at speeds close to TV-rate, paving the way for phase contrast imaging with temporal resolution for \textit{in-situ} experiments.
However, at significantly reduced dwell times the temporal response of the detector becomes important.
Several works have previously shown how scintillator-based detectors have a scintillation decay time which causes the duration of single electron events to last over $0.5$~\textmu s, sometimes as long as $1.5$~\textmu s~\cite{buban_high-resolution_2010,krause_effects_2016,sang_characterizing_2016,mittelberger_software_2018,mullarkey_development_2021,mullarkey_how_2023,peters_electron_2023}. 
When scanning at dwell times slower than around twice the scintillator decay time, the effects of this afterglow is negligible, but for faster scans, the image begins to suffer from streaking artefacts as a result of scintillator decay~\cite{buban_high-resolution_2010,mullarkey_how_2023}. 
A potential solution to this is to use low enough beam current to be able to count the individual electrons in the image, either in post-processing~\cite{ishikawa_quantitative_2014,mittelberger_software_2018}, or by live in-hardware digitisation of the signal from the detector photo-multiplier tube (PMT)~\cite{peters_electron_2023}. 
By digitising the signal from the detector in real-time, the streaking artefacts seen for analog images are removed and more information is retained in the fast scan direction.
\par

In this work, we combine for the first time ultra-fast scan coils~\cite{ishikawa_high_2020} and in-hardware digitisation of segmented scintillator-based detector~\cite{peters_electron_2023} to show that digitisation allows for much faster usable frame-rates and paves the way for low-dose phase characterisation, with potential use cases in beam-sensitive and low atomic number materials, as well as \textit{in-situ} experiments.
\par

\section{Low-dose approaches with DPC}
For beam-sensitive materials that either change characteristics or damage under electron beam irradiation, the possibilities for reduced dose is vital for precise characterisation. 
Collection efficiency can be thought of as the fraction of electrons that passed through the sample which are then used to generate signal. 
Whereas HAADF typically detects up to about $10$\% of the primary beam's electrons, an annular detector in the annular bright field (ABF) region may reach $\sim80$\% collection~\cite{klenov_contributions_2006}. 
Similarly, positioning a four-segment DPC detector in the ABF region yields an equally high collection efficiency~\cite{liang_optimizing_2023}.  
The high collection efficiency of detectors with an outer collection angle close to the semi-convergence angle ($\alpha$) allows for a significant reduction in the beam current used, and as such is excellent for imaging beam-sensitive materials.
\par 

The dose applied to the sample, in units of electrons per unit area, may be expressed as:
\begin{equation}\label{eq:doseeq}
    \mathrm{Dose} = \frac{I\cdot C \cdot \delta_t }{(\dd{x})^2}
\end{equation}
where $I$ is the beam current, $C$ the Coulomb number, $\delta_t$ the dwell time, and $\dd{x}$ the pixel width. 
Importantly, this only accounts for the exposure during the image capture, whereas the complete exposure also includes the sample exposure during beam flyback~\cite{mullarkey_using_2022}.
Still \autoref{eq:doseeq} clearly illustrates that reducing both beam current and dwell time is important to reach ultra-low dose conditions. 
When the $\delta_t$ reduction is limited for example for 4D-STEM imaging, the dose must be reduced by means of reducing the beam current or increasing the pixel width, since lowering the dwell time is not available.
Finally, dose fractionation by multi-frame imaging has been shown to significantly reduce beam damage to the sample~\cite{jones_smart_2015,jones_managing_2018}. 
Therefore, detectors with low enough read-out overhead to acquire many frames at a high enough frame rate for the sample to remain unchanged during a single frame capture also benefits beam sensitive samples. 
\par 

The phase shift imposed on the electron beam by the sample causes redistribution of the beam in the diffraction plane, which may be detected as a net displacement of the center-of-mass (COM) of the CBED pattern. 
This COM displacement is proportional to the gradient of the phase shift acquired by the beam which, in the absence of  magnetic fields in the specimen, is proportional to the electric field of the sample~\cite{dekkers_differential_1974,rose_nonstandard_1977}. 
With segmented detectors the COM in x-direction in detector space is estimated by:
\begin{equation}\label{eq:comx}
    \mathrm{COM}_x(\va*{r}) = \sum_i I_i(\va*{r}) \cdot \va*{k_x}_i
\end{equation}
where $\va*{r}$ is the position in real space, $\va*{k_x}_i$ is x-position of the center-of-mass of the $i$th detector segment~\cite{shibata_electric_2017}.
The expression for $\mathrm{COM}_y$ follows similarly. 
For a four-segmented detector these calculations simplify to subtracting opposing detector segments in x- and y-directions. 
The vector signal $\vb*{\mathrm{COM}}(\va*{r})$ is proportional to the projected electric field of the sample. 
It then follows from Gauss' law and the Poisson equation that the integrated signal, $\mathrm{iCOM}(\va*{r})$, is proportional to the projected electric potential of the sample~\cite{lazic_phase_2016}. 
This signal is typically calculated by integration in Fourier space,
\begin{equation}
    \mathcal{F}\qty{\mathrm{iCOM}(\va*{r})}(\va*{k}) = 
    \frac{\mathcal{F}\qty{\vb*{\mathrm{COM}}(\va*{r})}(\va*{k}) \cdot \va*{k}}{2\pi i k^2}
\end{equation}
where $\va*{k}$ is the position vector in reciprocal space~\cite{lazic_phase_2016}. 
We can therefore use this to retrieve the projected electric potential of our samples from the $\vb*{\mathrm{COM}}(\va*{r})$ signal obtained from \autoref{eq:comx}.
\par

\section{Optimising detector geometry for low-dose iCOM}
\begin{figure}
    \centering
    \includegraphics[width=.8\columnwidth]{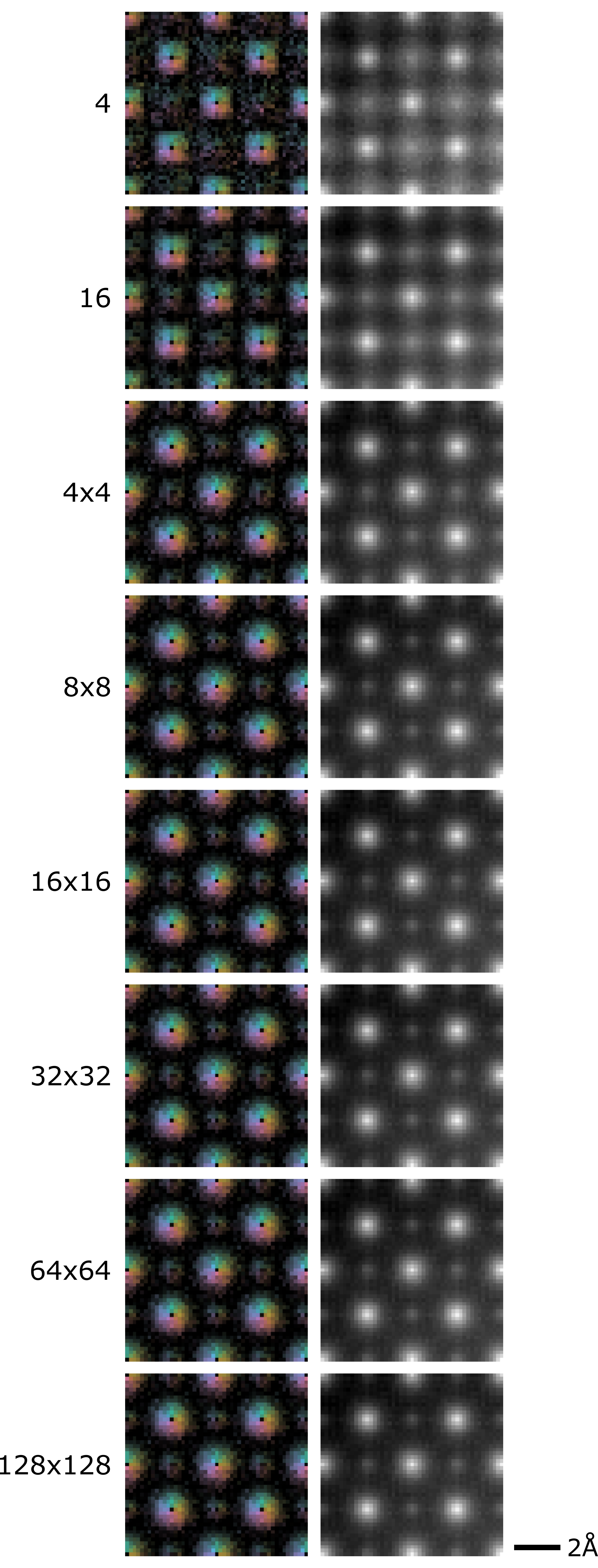}
    \caption{Finite dose (6241 e/Å$^2$) COM and iCOM images calculated from different virtual detector geometries.}
    \label{fig:finitedose_COM_iCOM}
\end{figure}
\begin{figure}[t]
    \centering
    \includegraphics[width=\columnwidth]{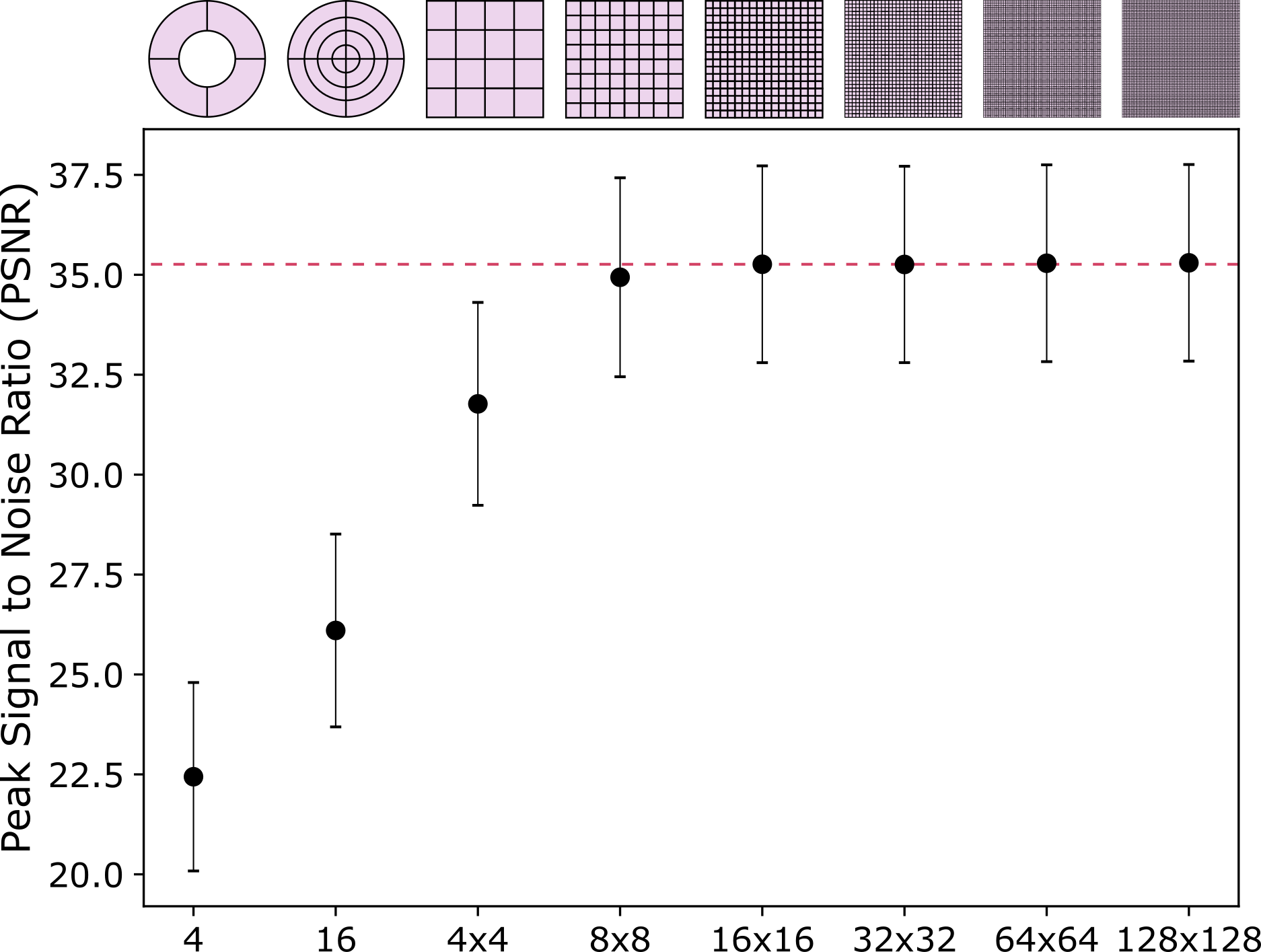}
    \caption{Peak signal-to-noise ratio in iCOM reconstructions from virtual detectors for a simulation of atomic resolution STO with 50 frozen phonon configurations and a finite dose of 6241 e/Å$^2$. The 4 denotes an annular detector with four segments and the 16 an annular detector with four rings of four segments each. Further, the rest denote pixelated array detectors. 
    The dashed line is at the PSNR of the 16x16 pixel detector, highlighting that even the 8x8 pixel detector performs just barely worse than the higher number pixel detectors.
    The error bars are estimated from the statistical error arising from the MSE between the infinite and finite dose images.}
    \label{fig:008_PSNR_exactvsnoisyiCOM_dose6241_61167}
\end{figure}
To investigate the fidelity of the COM signal and the reconstructed iCOM images from both segmented and pixelated detectors (with differing numbers of segments or pixels), a 4D dataset of a thin SrTiO$_3$ (STO) sample of $11.7$~nm (30 unit cells) thickness with 50 frozen phonon configurations was simulated using abTEM~\cite{madsen_abtem_2021}. 
To match the experimental data presented later, a $300$~keV beam energy and a $30$~mrad semi-convergence angle were used. 
The full 4D-STEM simulation (191x191 pixels) has a max collection angle of $60$~mrad (2$\alpha$).
Different detector geometries (masks) were  applied to the diffraction patterns and the resulting COM and iCOM images calculated using the python-based 4D-STEM data-handling software py4DSTEM~\cite{savitzky_py4dstem_2021}. 
To facilitate comparison with Yang et. al.~\cite{yang_efficient_2015} a finite dose of $6241$~e/Å$^2$ ($10$~C/cm$^2$) was applied representing Poisson statistical uncertainty.
A direct comparison of the resulting COM and iCOM images for eight different geometry detectors can be seen in \autoref{fig:finitedose_COM_iCOM}.
The segmented annular detectors, 4 and 16, with four and sixteen segments respectively show a slight four-fold symmetry in the reconstructed images due to the four-fold symmetry of the detectors~\cite{yang_efficient_2015,taplin_low_2016,li_integrated_2022}. 
This four-foldedness is eliminated for the pixelated arrays, even at the coarse pixelation of 4x4 pixels. 
\par 

The peak signal-to-noise ratio (PSNR) captures the ratio between the maximum possible power of a signal and the signal-disturbing noise. 
It is given as:
\begin{equation}
    \mathrm{PSNR} = 10 \log_{10}\qty(\frac{\mathrm{MAX}_i^2}{\mathrm{MSE}})
\end{equation}
where $\mathrm{MAX}_i$ is the maximum image value range in the infinite dose image and the MSE is the mean squared error between the infinite and finite dose images~\cite{yang_efficient_2015}.
Since we have both the full-field, infinite dose simulation and finite dose simulations we can define the PSNR as:
\begin{equation}
    \mathrm{PSNR} = 10 \log_{10}\qty(\frac{\mathrm{max}\qty(ref)^2}{(\overline{ref - expt})^2}).
\end{equation}
Here, $ref$ is the infinite dose reference image and $expt$ is the finite dose image~\cite{yang_efficient_2015}.
\autoref{fig:008_PSNR_exactvsnoisyiCOM_dose6241_61167} shows that the PSNR of the images in \autoref{fig:finitedose_COM_iCOM} reaches a plateau once the detector pixelation is 16x16 pixels. 
This calculation was repeated for several different finite doses, calculating both iCOM and magnitude of COM signals, and they all show the same result of minimal improvement of the PSNR for higher number of pixels in the detector beyond 16x16 (see Supplementary Information).
Since the improvement of iCOM phase retrieval is minimal for higher sampling of the CBED, the detector pixel count can be kept low to allow for faster scanning.
This is not to say that larger format cameras are not useful, in fact they are used widely for strain mapping~\cite{wang_autodisk_2022}, PACBED~\cite{lebeau_position_2010,hwang_nanoscale_2012}, crystallography characterisation with scanning electron (precession) diffraction (S(P)ED)~\cite{rauch_automated_2010,midgley_precession_2015,jeong_automated_2021}, and more exotic applications like multi-beam electron diffraction~\cite{hong_multibeam_2021} and patterned probe Bragg measurements~\cite{zeltmann_patterned_2020}. 
Rather, it is just that for their use in DPC and iCOM, the additional pixelation offers no advantage.
Still, regardless of whether a pixelated or segmented detector is used, DPC places strong limitations on the experimental parameters like sample thickness, bending and the microscope operator's ability to align the beam to zone axis. 
Because dynamical scattering and zone axis changes can produce a DPC signal indistinguishable from the DPC signal expected from an electric or magnetic field, careful considerations must be given to both sample preparation and the setup of the experiment~\cite{maclaren_origin_2015,burger_influence_2020}.
\par

4D-STEM detection affords the opportunity to detect all the electrons from $0-\alpha_{max}$, where $\alpha_{max}$ is the outer collection angle set by the camera length, often at high resolution. 
While this offers very precise estimates for the COM signal, there are several other works that suggest losing the center of the CBED, as we do when using an annular segmented detector, causes minor loss to the final iCOM signal~\cite{muller_atomic_2014,shibata_differential_2012,majert_high-resolution_2015,shibata_electric_2017,seifer_flexible_2021,kohl_optimized_2023}. 
A similar result has been shown for ptychography~\cite{song_hollow_2018}.
It has previously been shown that the PSNR of single side-band (SSB) ptychographic reconstruction of an isolated carbon atom increases very modestly for higher detector pixelation than 16x16 pixels~\cite{yang_efficient_2015}. 
Even segmented annular detectors of 4 and 16 segments can provide comparable ptychographic reconstructions.
This further demonstrates how increasing pixelation of the detector is not necessarily better for phase contrast imaging.
We may therefore reap the benefits of fewer element detectors (be they segments or pixels), to enable fast scanning and multi-frame to image with high spatial and temporal resolution.
It should however be mentioned that low-dose ptychography is achievable and routinely performed using iterative ptychographic reconstruction algorithms~\cite{rodenburg_phase_2004,maiden_improved_2009,song_atomic_2019,zhou_low-dose_2020,li_4d-stem_2022}. 
These experiments often use highly defocused probes allowing large sample step sizes, which also improves the achievable temporal resolution. 
Additionally, more advanced methods like multi-slice ptychography and tilt-corrected BF-STEM are becoming more and more popular and showing promising results~\cite{maiden_ptychographic_2012,varnavides_simultaneous_2023,ribet_uncovering_2024,oleary_three-dimensional_2024}. 
For such techniques, increased pixelation of the detector beyond 16x16 is advantageous.
\par

\section{Experimental digitisation}
\begin{figure}[htb]
    \centering
    \includegraphics[width=\columnwidth]{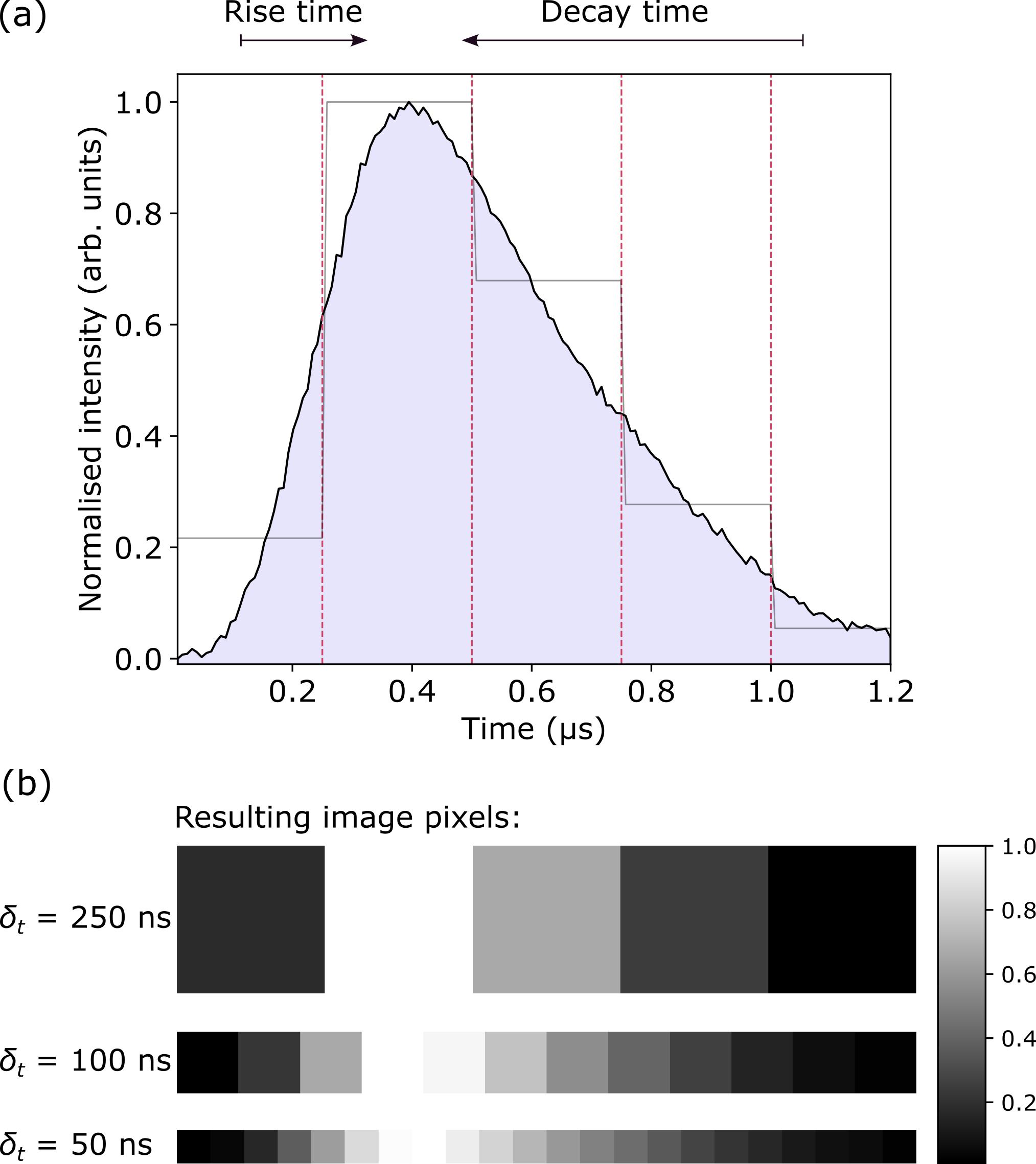}
    \caption{Single pulse from the SAAF detector with the 10-90\% rise time and 90-10\% decay time indicated. 
    The vertical, dotted lines are the pixel boundaries for a $250$~ns dwell time, and the stepped gray line indicates the normalised pixel intensity for each pixel with this dwell time. 
    The (normalised) resulting image pixels are shown in the diagram below, for dwell times $\delta_t = 250, 100, 50$~ns respectively.
    }
    \label{fig:Singlepulse_with_dt0.25us_240222}
\end{figure}
Using a probe-corrected JEOL Grand ARM operated at 300kV, single crystals of [100] oriented SrTiO$_3$ were imaged with a 5.2 pA beam current. Several fast dwell times in the range 50 ns to 1 µs were collected with both with analog and digital signal recording. 
The microscope used is equipped with fast low-inductance scan coils allowing dwell times as short as $50$~ns to be reached~\cite{ishikawa_high_2020}. 
For the DPC imaging we used a segmented annular all-field (SAAF) detector, which has 16 segments spread over four concentric rings with four segments each~\cite{shibata_new_2010}. 
Performing the in-hardware digitisation has to be done on a per-segment basis and thus one channel per segment must be provided.
For the experiments presented here, a four channel Pulse digitiser produced by turboTEM was used and the second ring from the center was chosen for the DPC imaging (see Supplementary Information). 
Digitisation of 8, 12, or 16 channels is possible by using additional Pulse digitisers in parallel so long as the scan unit can record their outputs (in this case a Gatan Digiscan 3 accepting four simultaneous channels was used).
The outer collection angle of the detector annulus used was $32$~mrad and the inner collection angle $16$~mrad, for an $\alpha = 30$~mrad beam. 
A Pulse digitiser is easily retrofittable to any STEM system with BNC cables, and the main demand is that the scan generator accepts the number of channels desirable -- preferably directly as digital inputs.
Both analog and digitised signals may be recorded simultaneous by forking the signal physically with a T-piece, provided that the number of input channels is sufficient.
In order to facilitate multiple, parallel signal digitisation and multi-frame imaging, both analog and digital signal inputs are beneficial as well as a frame clock to aid multi-framing. 
A screenshot of the software for controlling the Pulse digitiser can be seen in the Supplementary Information.
\par 

Scintillator-based detectors have been shown to have quite significant electron detection event decay time, causing streaking artefacts in the images~\cite{mullarkey_how_2023,mittelberger_software_2018,sang_characterizing_2016,krause_effects_2016}.
A single electron event on the SAAF detector can be seen in \autoref{fig:Singlepulse_with_dt0.25us_240222}(a), with dotted lines indicating pixel boundaries for a dwell time of $250$~ns. 
The full duration of the event is a combination of the time it takes for the scintillator to respond to the electron impact, the PMT to amplify the resulting light signal, and the read-out through the cables to the recording hardware. 
Additionally, scintillators often suffer from after-glow, meaning that they keep scintillating slightly for an extended time after exposure -- even when the electron exposure is removed~\cite{mullarkey_development_2021}.
When the duration of any single electron event exceeds the pixel dwell time, the electron which was physically incident at the detector at a specific moment leads to signal intensity also for the following pixels. 
This causes streaking in the images, which may be seen in \autoref{fig:Singlepulse_with_dt0.25us_240222}(b), and as a result information about the sample is degraded in the fast scan direction~\cite{mullarkey_how_2023,buban_high-resolution_2010}. 
A consequence of the event duration of over a microsecond on this detector is that dwell times lower than that should be avoided to ensure the information integrity in the analog images. 
However, if the electron events are digitised using hardware for live PMT signal digitisation, lower dwell times become viable again~\cite{peters_electron_2023}. 
\par

\begin{figure*}[htb!]
    \centering
    \includegraphics[width=\textwidth]{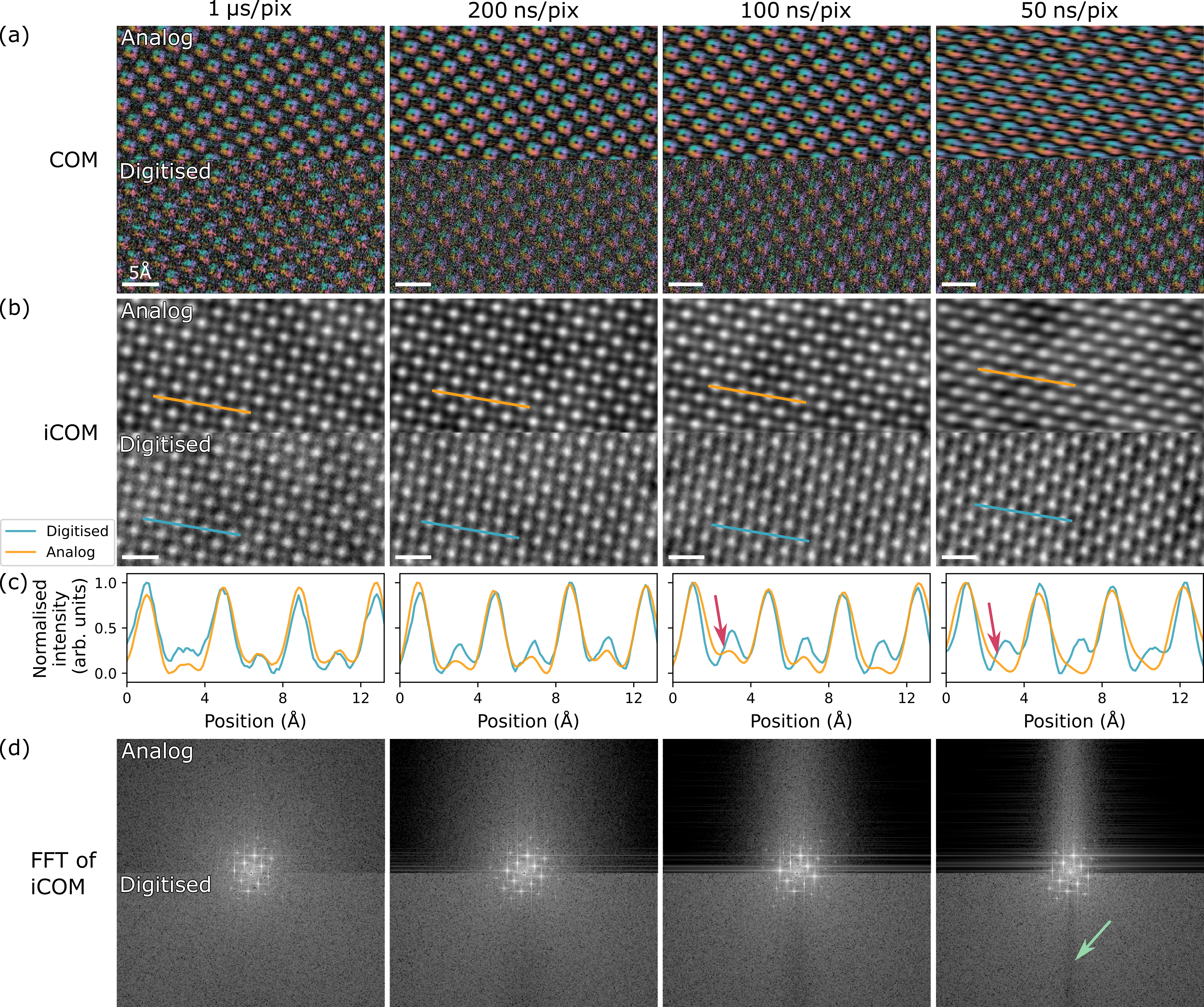}
    \caption{(a,b) Experimental COM and iCOM images, analog (top) and digitised (bottom), of STO acquired with beam current $I = 5.2$~pA and increasingly (left to right) lowered dwell times. 
    The resolution in the analog images is severely reduced in the fast scan (horisontal) direction due to image streaking. 
    (c) Line profiles from the lines indicated in (b), showing how the oxygen columns become increasingly unresolvable with analog DPC at the fastest scan speeds. 
    (d) FFTs of iCOM images in (b) show how information is lost in the fast scan direction due to streaking.}
    \label{fig:Experimental_analog_digitised_comparison_231005_SrTiO3_DPC}
\end{figure*}

The in-hardware digitisation calculates the gradient of the PMT signal and thresholds this to count the electron events. 
This results in a digitised signal which has a true zero noise floor, and where each electron is detected in the correct pixel only and all have the same intensity of unity. Digitising the signal from scintillator detectors has previously been demonstrated to significantly improve signal to noise ratio in HAADF images~\cite{mullarkey_development_2021,peters_electron_2023}.
Since the digitised pulse appears in the correct temporal position in the PMT stream it also appears in the correct spatial position in the image, without affecting subsequent pixels. As a result, all streaking artefacts are removed. 
Incidentally, there is some unavoidable few nanoseconds progressing time required by the digitisation hardware, but this manifests as a  uniform apparent translation of the sample to the right by some constant fraction of a pixel width.
Since scintillator detector surfaces are typically inhomogenous, the electrons may be detected at different intensity depending on where on the detector surface they hit~\cite{macarthur_how_2014,sang_characterizing_2016}.
Consequently, digitisation of the analog signal ensures a more homogeneous detector response, and yields an image that is quantitative, in number of electrons~\cite{macarthur_how_2014,mullarkey_how_2023}.
This ensures that detector inhomogeneity does not affect the COM calculations, and that denoising algorithms developed for quantised signals~\cite{kusumi_new_2024} may be used on the individual segment images before computing the COMs (although no denoising was used in this study).
As a final advantage, having digitised signals makes quantitative comparison with simulations easier.
\par 

\autoref{fig:Experimental_analog_digitised_comparison_231005_SrTiO3_DPC} shows how streaking artefacts severely reduce information transfer in the fast scan direction for the analog images.
The subtraction of one weak signal from another can be intrinsically noisy. In DPC, subtracting opposing detector segment signals can also lead to the digital COM images appearing very noisy. 
This differs from the analog detector response which intrinsically performs a form of low pass filtering, yielding visually less noisy images.
However, it is important to note that although the smoother images may be more visually appealing, this smoothing is the result of spatial resolution loss. By contrast, the iCOM reconstruction from the digitised signals show a very minimal information degradation, even at the fastest scan speed similar to the results for digital ADF imaging~\cite{mullarkey_development_2021,peters_electron_2023}.
Where the streaking cause the oxygen columns to be lost in the faster scan analog iCOM images, see line profiles in \autoref{fig:Experimental_analog_digitised_comparison_231005_SrTiO3_DPC}, the digitised iCOM retain these columns very clearly.
Notice especially how the background of the fast Fourier transforms (FFTs) are much flatter for the digitised iCOM images as compared with the analog iCOM images. 
This indicates that the information transfer is also much more isotropically reliable when digitising the PMT signal, and thus that the digitised iCOM images are more isotropically reliable.
\par

For a STEM detector spanning the ADF region, a highly scattering point on the specimen will increase the arrival rate of electrons to the detector. Such mass-thickness type contrast often follows a $Z^n$ relationship where $n$ is around 1.7 depending on collection angle.
Weakly scattering materials on the other hand are often studied with DPC due to the higher sensitivity for lower atomic numbers.
Placing the DPC detector in the ABF region necessarily increases the arrival rate of electrons, relative to ADF, since the collection angles are within the semi-convergence angle. 
Although this aids collection efficiency, like all discrete counting processes, the arrival of two events approximately simultaneously may cause coincidence losses; a risk which increases dramatically at high beam currents.
Using a significantly lowered beam current is then a convenient solution both to the need to avoid sample beam-damage, and avoid coincidence loss.
Since coincidence loss is a process between adjoining detection events in \textit{time} and not in space, these are only a factor along the fast-scan direction (and not between successive scan lines). 
This manifests in the Fourier transform (FT) as the formation of a dark band aligned perpendicular to the fast-scan direction. 
Since the fast-scan direction is presented horizontally by convention, this dark band runs vertically through the FT (see arrow in~\autoref{fig:Experimental_analog_digitised_comparison_231005_SrTiO3_DPC}(d)).
The appearance of the faint vertical dark line in the FFT of the digital iCOMs suggests that at this beam current ($I = 5.2$~pA) some electron events are lost to coincidence loss. 
In the slower scans the vertical dark band becomes broader since the coincidence loss persists at the same temporal domain but this becomes a higher spatial frequency in the image domain.
By simulation, we have determined that the minimum separation for digitisation of two consecutive electron events for this specific detector is $259 \pm 26$~ns (for details see Supplementary Information).
Importantly, this depends heavily on the beam current and the temporal response of the detector.
For the beam current and detector used here this translates to loss of just under one electron per pixel for $\delta_t = 250$~ns, and approximately one every five pixels for $\delta_t = 50$~ns.

\section{Time-resolution in multi-frame series}
\begin{figure*}[h!]
    \centering
    \footnotesize
    \includegraphics[width=\textwidth]{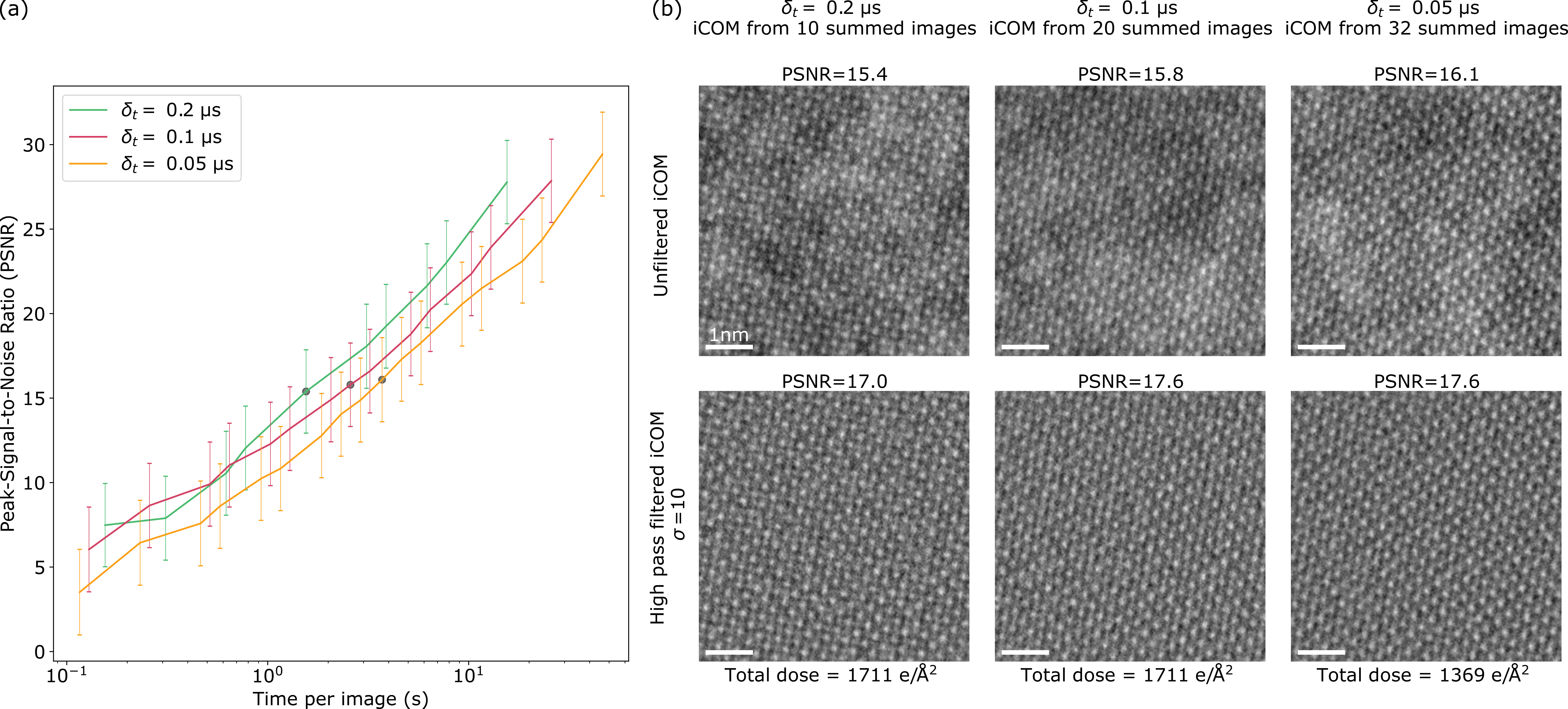}
    \normalsize
    \caption{PSNR in iCOM images for various summing of multi-frame stacks. 
    Time per image is then given as the combined time it took to acquire all the frames, e.g. for 20 summed frames the time per image is $20 \cdot T_\mathrm{Frame}$. 
    The images on the right are the result of summing 10, 20 and 32 frames respectively for the three different dwell times $\delta_t = 0.2, 0.1, 0.05$~\textmu s. The top images are raw summed images and the bottom images are high-pass filtered.}
    \label{fig:PSNRvstimeperimage_w_exampleimages}
\end{figure*}
Reducing the beam current and using very low dwell times reduces the signal recorded in a single frame acquisition. In extreme cases, the image may contain so little signal that features are no longer distinguishable and phase reconstruction may fail.
A useful signal level may be regained by scanning, drift correcting, and stacking many subsequent frames~\cite{jones_smart_2015,ishikawa_atomic-resolution_2021}.
In this way, the sample exposure to the beam is spread out over multiple passes of shorter exposures, reducing sample damage and enhancing signal integrity~\cite{jones_optimising_2017}.  
As would be expected, the PSNR increases with increasing number of summed frames as shown in \autoref{fig:PSNRvstimeperimage_w_exampleimages}.
As an example, achieving a time resolution of $4$ seconds per output frame (at 512x512 image pixels) may be realised using $20$ individual digital-DPC scan frames, each with $\delta_t = 50$~ns.
Doing so yields an image with a PSNR of $16$ without any denoising applied.
\par

The iCOM phase reconstruction generally suffers from low-frequency artefacts, which are commonly filtered out using a high-pass filter~\cite{yucelen_phase_2018,lazic_single-particle_2022}. 
Applying a high-pass filter to the images in \autoref{fig:PSNRvstimeperimage_w_exampleimages} increases the PSNR by removing the low-frequency artefacts that arise from the phase reconstruction. 
This was achieved by subtracting a Gaussian filtered ($\sigma = 10$) version of the image. 
For consistency, the same high-pass filter was used for all images (see Supplementary Information for the difference images).
That means that, depending on the necessary PSNR for the desired information to be extracted, time resolution of a few seconds may be realised with multi-frame, digital DPC imaging. 
Compared to 4D-STEM-based methods, where dwell times must typically be larger than $100$~\textmu s and the time to acquire a single frame often reaches a few minutes, or sacrifice sampling or field-of-view, this is a huge improvement.
\par 

In the graph in \autoref{fig:PSNRvstimeperimage_w_exampleimages}a we can see how the PSNR increases with increasing "time per image", which indicates that the more frames are summed the higher the PSNR in the final analysable image. 
It is also evident that for a given "time per image", the PSNR is generally higher if the frames are captured with a slower dwell time.
While this might seem somewhat discouraging for adopting faster scan speeds, a major factor in this is the fly-back time. 
The faster the scan, the more frames are required to get enough signal for an analysable image, and the more times the beam must fly back. 
As a result, the time lost to line and frame fly-back is more significant for the faster scan and the temporal resolution achievable from the multi-frame stack may be worsened.
In summary, it is vital in these experiments to balance the need for low beam current for viable digitisation, fast scanning for dose fractionation and temporal resolution, and fly-back time considerations. 
As a final note, with faster scintillators the practicable beam current is increased and with lower-hysteresis scan coils or fly-back compensation~\cite{mullarkey_using_2022} the temporal resolution may be increased. 
Both of which, although beyond the scope of this work, may in the future increase the fidelity and practicability of digitised DPC and improve the temporal resolution for \textit{in-situ} phase characterisation.

\section{Conclusions}
Due to camera read-out overhead, the relatively slow dwell times required for 4D-STEM can introduce unacceptable image artefacts arising from sample drift or environmental instability.
Additionally, the many-GB datasets generated by 4D-STEM cameras may be cumbersome to acquire, store and process.
Scintillator-based segmented detectors exhibit negligible read-out overhead and far smaller dataset sizes, but where analog read-out speeds approach the scintillator decay-time other unacceptable artefacts arise.
Smearing of the atomic columns in space, due to signal streaking in time, make the precise characterisation of subtle atomic column displacements virtually impossible.
The demand for increased dwell times in both these cases stands in the way of ultra low-dose phase contrast imaging. 
\par

In this work, by in-hardware digitisation of the signal from a scintillator-based segmented detector, we have for the first time demonstrated phase imaging at massively improved speeds using a dwell time of $50$~ns. 
We show that the digitised images, acquired at a relatively low beam current ($I = 5.2$~pA) suffer minimally from coincidence loss, even with the detector under the direct forward beam. 
Multi-frame bursts of ultra-fast acquired COM images are used to delay sample damage, accumulate sufficient signal for high-quality iCOM phase reconstruction, and retain temporal resolution for future \textit{in-situ} studies in the STEM.

\par 

\section*{Acknowledgements}
The authors would like to acknowledge the Advanced Microscopy Laboratory (AML) at the Centre for Research on Adaptive Nanostructures and Nanodevices (CRANN) and the Advanced Materials and BioEngineering Research (AMBER) Network for financial and infrastructural support for this work. JMB and JJJP acknowledge SFI grant number 19/FFP/6813, RI and NS acknowledge JST FOREST (Grant No. JPMJFR2033), JST ERATO (Grant No. JPMJER2202), and JSPS KAKENHI (Grant No. JP19H05788, 20H05659, JP22H04960, JP24H00373), and LJ acknowledges Royal Society and SFI grant number URF/RI/191637 and 12/RC/2278\_P2.

\section*{Declaration of Interests}
LJ is the creator of the SmartAlign image registration software used here and made available commercially by HREM Research (Tokyo). LJ and JJPP are co-founders and equity holders of turboTEM Ltd. (Dublin) who manufacture the `Pulse' hardware signal digitiser used in this work.

\printbibliography

\clearpage
\onecolumn

\appendix
\pdfoutput=1
\renewcommand{\thefigure}{S\arabic{figure}}
\renewcommand{\thetable}{S\arabic{table}}
\renewcommand{\thesubsection}{S\arabic{subsection}}
\titleformat{\section}{\LARGE}{\thesection}{1em}{}

\setcounter{figure}{0}

\section*{Supplementary Information}

\subsection{4D-STEM detectors}
Summarised in \autoref{tab:detectors_information} are numbers representing the dynamic ranges of the commercially available 4D-STEM detectors displayed in Figure 1 of the paper. 
While it is inadvisable to compare the dynamic range of detectors directly using either of the given numbers, they are important to take into account because certain experiments may require data with a higher dynamic range. 
The bit depth gives information about how many bits is used to read out from the detector the information contained in each frame.
Note that bit depth does not necessarily tell the full story of the dynamic range of a detector since bits may be used to store other information than the intensity values directly.
Additionally, \autoref{tab:detectors_information} gives the highest beam current a detector can withstand without saturating in electrons per pixel per second. 
This is a physical property of the detector, however it can not be used to compare detectors directly because it does not account for their differing number of pixels. 
\begin{table}[h]
    \centering
    \caption{Detector types and dynamic range, given in bit depth and maximum incident beam current before saturation in e$^-$/pixel/s.
    }
    \label{tab:detectors_information}
    \begin{tabular}{l l l l}
        Detector & Type & Bit depth & Maximum count rate (e$^-$/pixel/s) \\
        \hline \\[-3mm]
        EMPAD       & PAD   & 30    & $1.2\cdot10^7$\\
        EMPAD2      & PAD   & 32    & $10^9$ \\
        Arina       & PAD   & 8-32\tablefootnote{8-32 image bit depth, 12 read-out bit depth.} & $10^8$ \\
        Ela         & PAD   & 8-16\tablefootnote{Speed given in the plot for the Dectris Ela is for the 8 bit read-out.}  & $6.2\cdot10^6$ \\
        UltraScan   & CCD   & 16    & - \\
        Orius       & CCD   & 14    & - \\
        OneView     & MAPS  & 12    & $2.25\cdot10^5$  \\
        Rio         & Scintillator CMOS & 12 & $6\cdot10^5$ \\
        K2          & MAPS  & 32    & $10^*$ \\ 
        K3          & MAPS  & 32\tablefootnote{The K3 reads out at 32 bit in linear accumulation mode, and 8 bit in counting mode.} & $40^*$ \\
        Metro       & MAPS  & 32    & $80^*$ \\
        ClearView   & MAPS  & 12    & $3\cdot10^6$  \\
        Medipix 1   & PAD   & 15    & - \\
        MerlinEM    & PAD   & 1-24\tablefootnote{For the Medipix/MerlinEM detectors that support a large range of bit depths the speed given in the plot is for 12 bit, but the acquisition speed may be increased by using a lower bit depth.}  & $10^6$ \\
        Medipix 4   & PAD   & 1-24 & - \\
        pnCCD       & CCD   & 16    & $4\cdot10^5$ \\
        Celeritas   & MAPS  & -     & $8.7\cdot10^5$ \\
        DE-16       & MAPS  & -     & $10^5$ \\
    \end{tabular}
    \begin{flushleft}
        \footnotesize$^*$ Recommended maximum dose rate to ensure detector linearity.\normalsize
    \end{flushleft}
\end{table}

\clearpage
\subsection{COM with varying detector geometries}
\begin{figure*}[!ht]
    \centering
    \includegraphics[width=.9\textwidth]{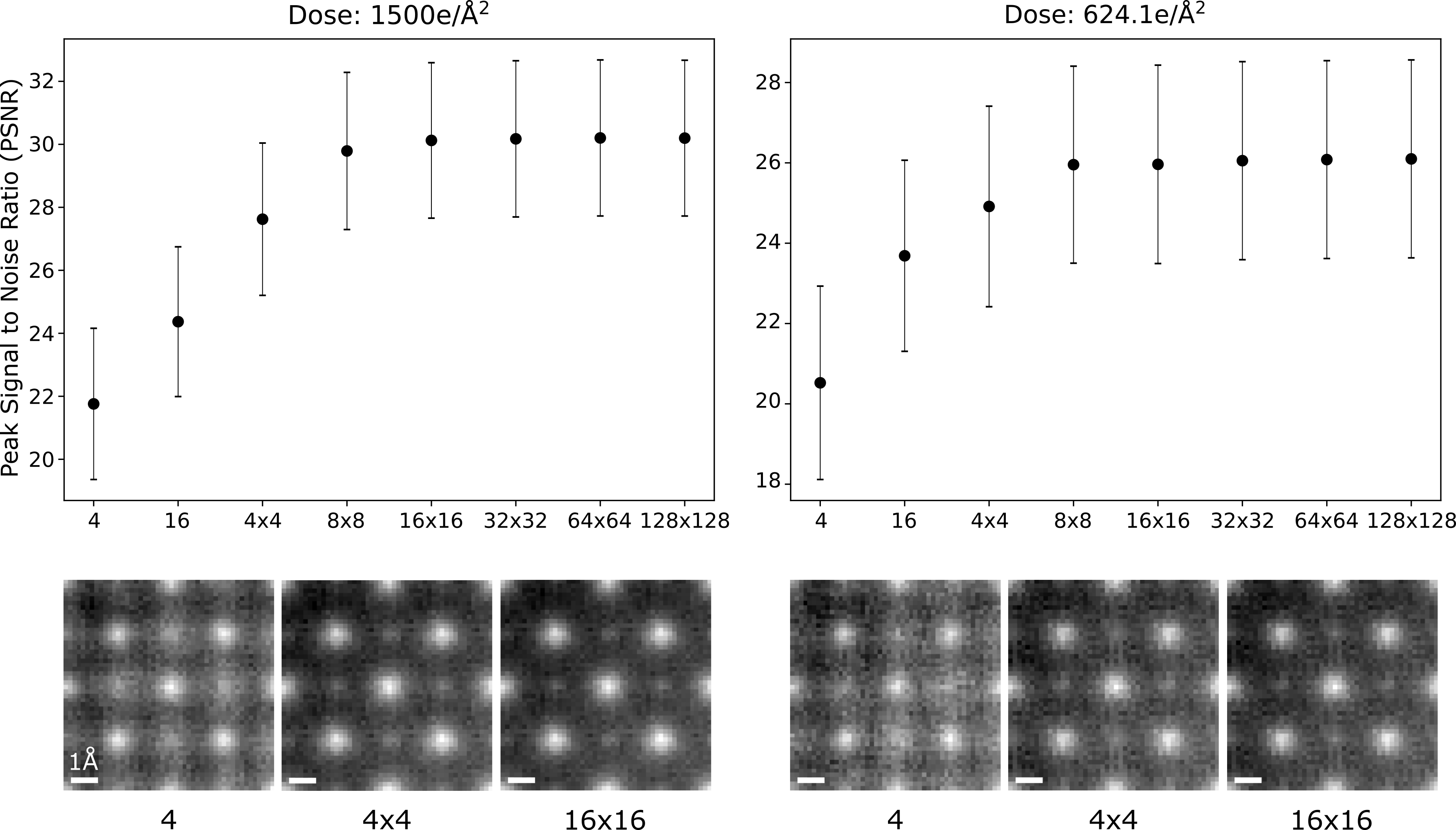}
    \captionof{figure}{Peak Signal to Noise Ratio (PSNR) in simulated finite dose iCOM images. Cut-out of the top left corner of the reconstructed images for detector geometries annular four-segmented, 4x4 pixel array, and 16x16 pixel array are shown beneath.}
    \label{fig:SI_PSNR_iCOM_finitedoses1500_624}
\end{figure*}

\begin{figure*}[!ht]
    \centering
    \includegraphics[width=.9\textwidth]{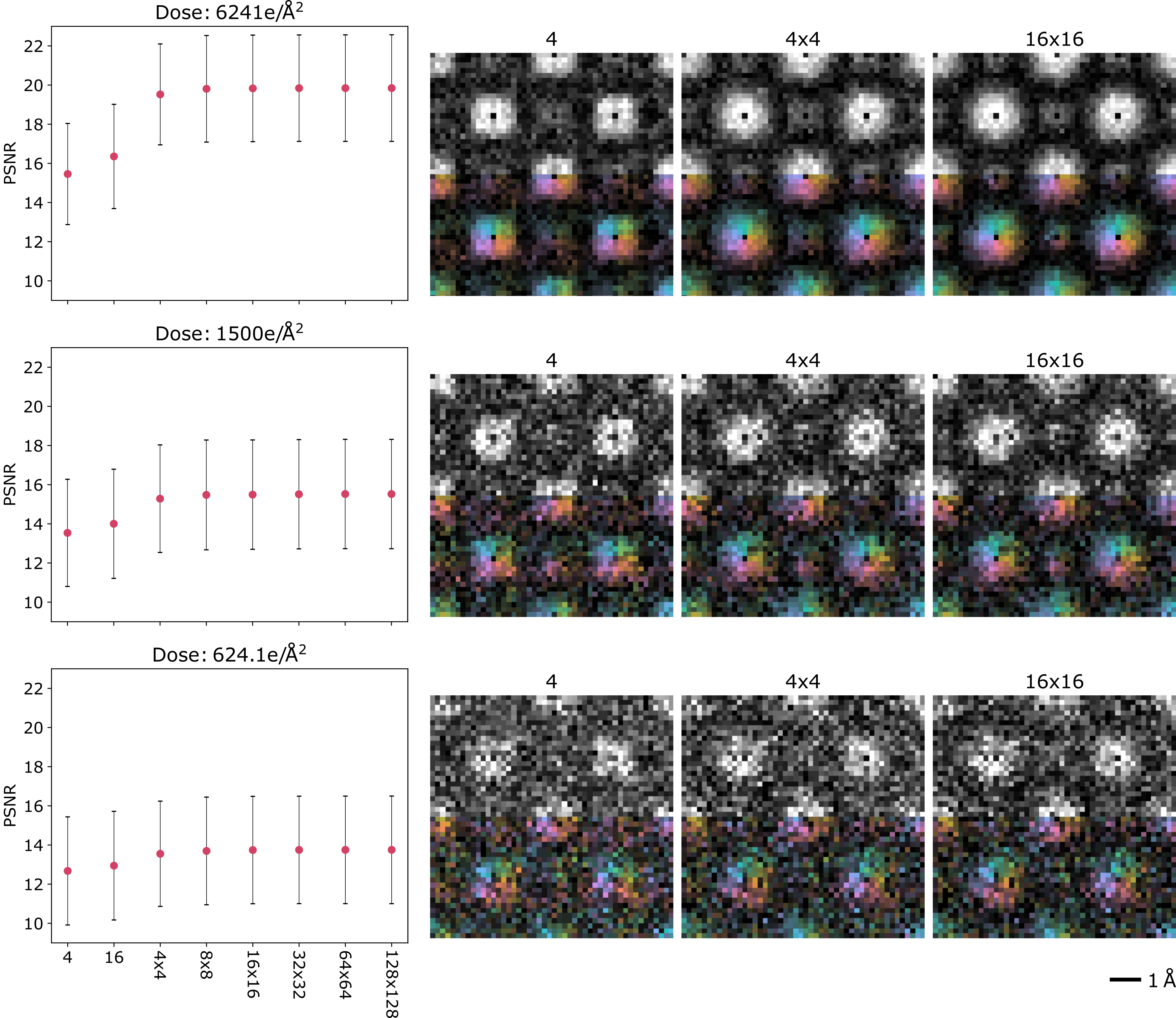}
    \caption{PSNR in simulated \textbar COM\textbar~images for finite doses $6241$, $1500$, and $624.1$ e/Å$^2$, plotted with the same y-axis in order to clearly show how the PSNR goes down with decreasing dose. The reference image is the infinite dose \textbar COM\textbar~image as calculated from the full 4D-STEM }
    \label{fig:SI_PSNR_COMmag_finitedoses6241_1500_624}
\end{figure*}
\clearpage

\begin{figure*}[!ht]
    \centering
    \includegraphics[width=.5\textwidth]{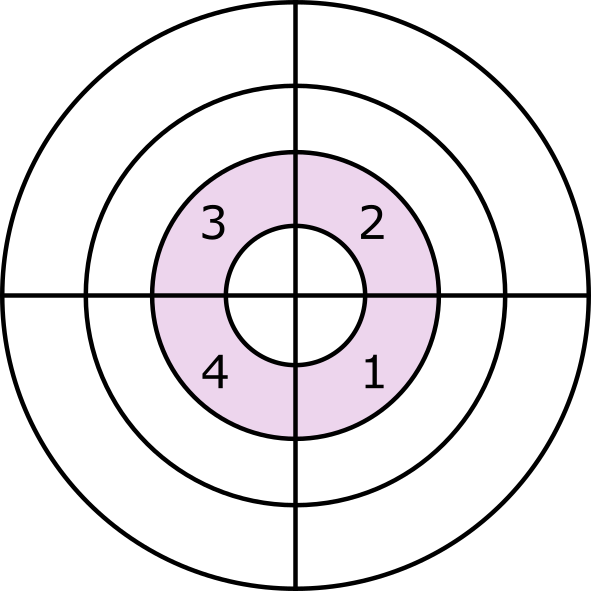}
    \caption{Illustration of the geometry of the segmented annular all field (SAAF) detector. 
    The second ring (highlighted) was used for both the analog and digitised imaging in this study.}
    \label{fig:SAAFillustration}
\end{figure*}

\clearpage
\subsection{Experimental digitisation}
\begin{figure*}[!ht]
    \centering
    \includegraphics[width=.9\textwidth]{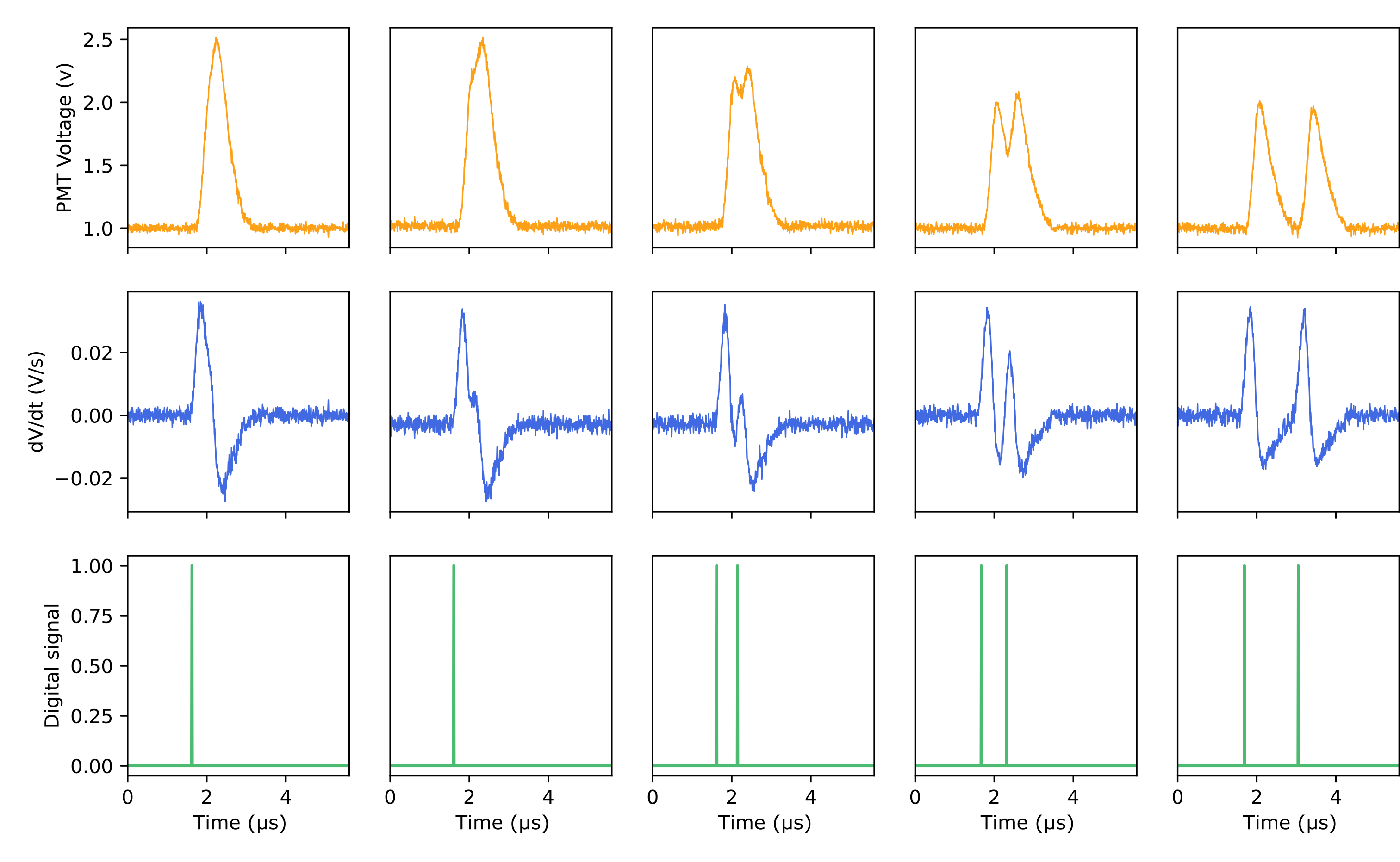}
    \caption{Illustration of electron detection events with increasing separation in time eventually allowing them to be digitised as two separate events. 
    The statistically acceptable separation time for two events to be separated was determined by simulating two events at increasing separation and recording when they were detected as two digitised events. 
    The average separation of $259 \pm 26$~ns was obtained from $200$ such simulations.}
    \label{fig:AcceptableEventSeparation}
\end{figure*}

\begin{figure*}[!ht]
    \centering
    \includegraphics[width=.9\textwidth]{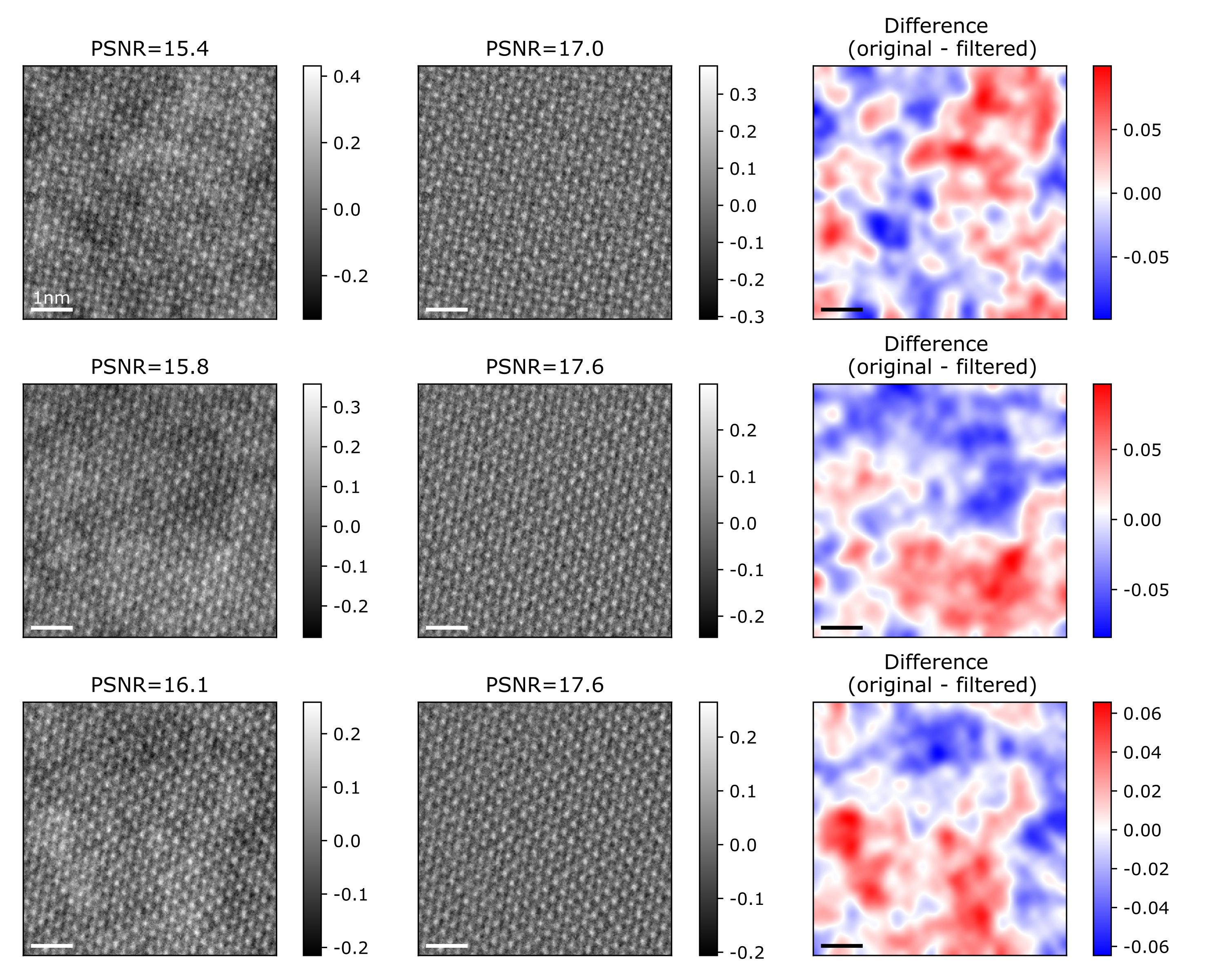}
    \caption{Unfiltered and high-pass filtered experimental images and their difference. 
    The high-pass filtered image was computed by subtracting a low-pass filtered version of the unfiltered image (Gaussian blur with $\sigma = 10$).}
    \label{fig:Difference_original-filtered_sigma10}
\end{figure*}

\begin{figure*}[!ht]
    \centering
    \includegraphics[width=\textwidth]{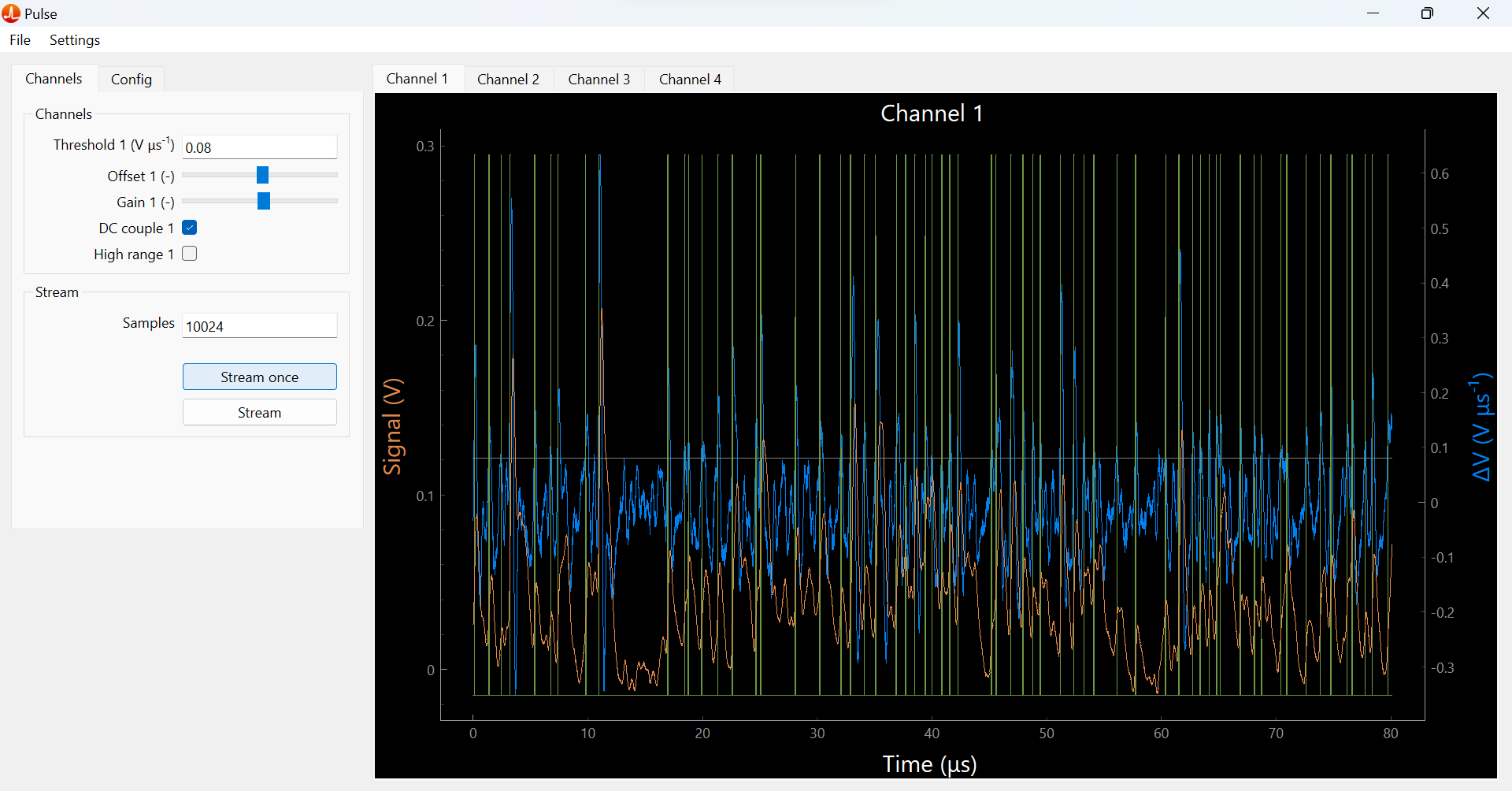}
    \caption{Screenshot of the software controlling the Pulse digitiser unit. 
    The display shows the raw data stream (in orange), the gradient (in blue), the threshold (horisontal gray line), and the digitised events (in green) for Channel 1. 
    The sidebar shows the settings for this channel.}
    \label{fig:Pulse_software}
\end{figure*}

\end{document}